\documentclass[aps,prl,twocolumn,superscriptaddress]{revtex4-1}

\usepackage[T1]{fontenc}
\usepackage[latin9]{inputenc}
\usepackage{amsmath}
\usepackage{amssymb}
\usepackage{graphicx}
\usepackage{hyperref}
\usepackage{natbib}
\usepackage{esint}
\usepackage{times}
\usepackage{helvet}
\usepackage{braket}
\usepackage{bbm}
\usepackage{xcolor}
\usepackage{enumitem}
\usepackage[toc,page,titletoc]{appendix}
 
\makeatletter
\newcommand{\JILA}{JILA, NIST and Department of Physics, University of Colorado, Boulder, CO 80309, USA}
\newcommand{\CTQM}{Center for Theory of Quantum Matter, University of Colorado, Boulder, CO 80309, USA}

\begin{document}

\title{Generating multipartite spin states with fermionic atoms in a driven optical lattice}
\date{\today}

\author{Mikhail Mamaev}
\email{mikhail.mamaev@colorado.edu}
\affiliation{\JILA}
\affiliation{\CTQM}
\author{Ana Maria Rey}
\affiliation{\JILA}
\affiliation{\CTQM}

\begin{abstract}
{We propose a protocol for generating generalized GHZ states using ultracold fermions in 3D optical lattices or optical tweezer arrays. The protocol uses the interplay between laser driving, onsite interactions and external trapping confinement to enforce energetic spin- and position-dependent constraints on the atomic motion. These constraints allow us to transform a local superposition into a GHZ state through a stepwise protocol that flips one site at a time. The protocol requires no site-resolved drives or spin-dependent potentials, exhibits robustness to slow global laser phase drift, and naturally makes use of the harmonic trap that would normally cause difficulties for entanglement-generating protocols in optical lattices. We also discuss an improved protocol that can compensate for holes in the loadout at the cost of increased generation time. The state can immediately be used for quantum-enhanced metrology in 3D optical lattice clocks, opening a window to push the sensitivity of state-of-the-art sensors beyond the standard quantum limit.}
\end{abstract}
\maketitle

%%%%%%
\textit{Introduction. }
%%%%%%
Creating useful entanglement is one of the most important goals in modern quantum research. In recent years, there has been significant effort towards generating multi-body entangled states, which exhibit massive utility for quantum computation, simulation and metrology. For the latter application of metrology, an $N$-body fully entangled state can yield sensitivity improvement by a factor of $\sqrt{N}$ compared to experiments using unentangled atoms or modes~\cite{Degen2017}. Such gains in precision are relevant for real-world applications such as time-keeping, magnetometry and navigation, and for fundamental science including searches for dark matter and physics beyond the Standard Model~\cite{DeMille990}.

While there has been progress on many-body entanglement generation in many fields, one of the most promising platforms is ultracold atoms. A variety of entangled states have been proposed and/or experimentally realized with such systems, including spin-squeezed states~\cite{ma2011squeezing}, W-states~\cite{haas2014wstate}, and in particular generalized GHZ (Greenberger-Horne-Zeilinger) states using trapped ions~\cite{Sackett2000,monz2011catState,leibfried2005catGenerationIons,Friis2018} or Rydberg atoms in optical tweezers~\cite{omran2019catState}. However, the difficulty of combining single-site resolution with scalability has limited the fidelity and size of the states thus far, especially in systems where they can be directly used for metrological purposes.

In this work, we propose a method for generating $N$-particle GHZ states (also called spin cat states) using ultracold fermionic atoms loaded into a 3D optical lattice. Our protocol uses onsite repulsive interactions, spin-orbit coupled (SOC) laser driving~\cite{kolkowitz2017soc,livi2016soc,Bromley2018}, and the harmonic trapping potential naturally generated by the curvature of the lattice beams. While we focus on 3D lattices, the setup may also be realized in optical tweezer arrays with an AC-Stark shift gradient to emulate the trap. We describe a step-by-step generation of entanglement by creating an initially-local superposition, and spatially changing one of its components while leaving the other component untouched due to energetic constraints.

\begin{figure}
\centering
\includegraphics[width=1\linewidth]{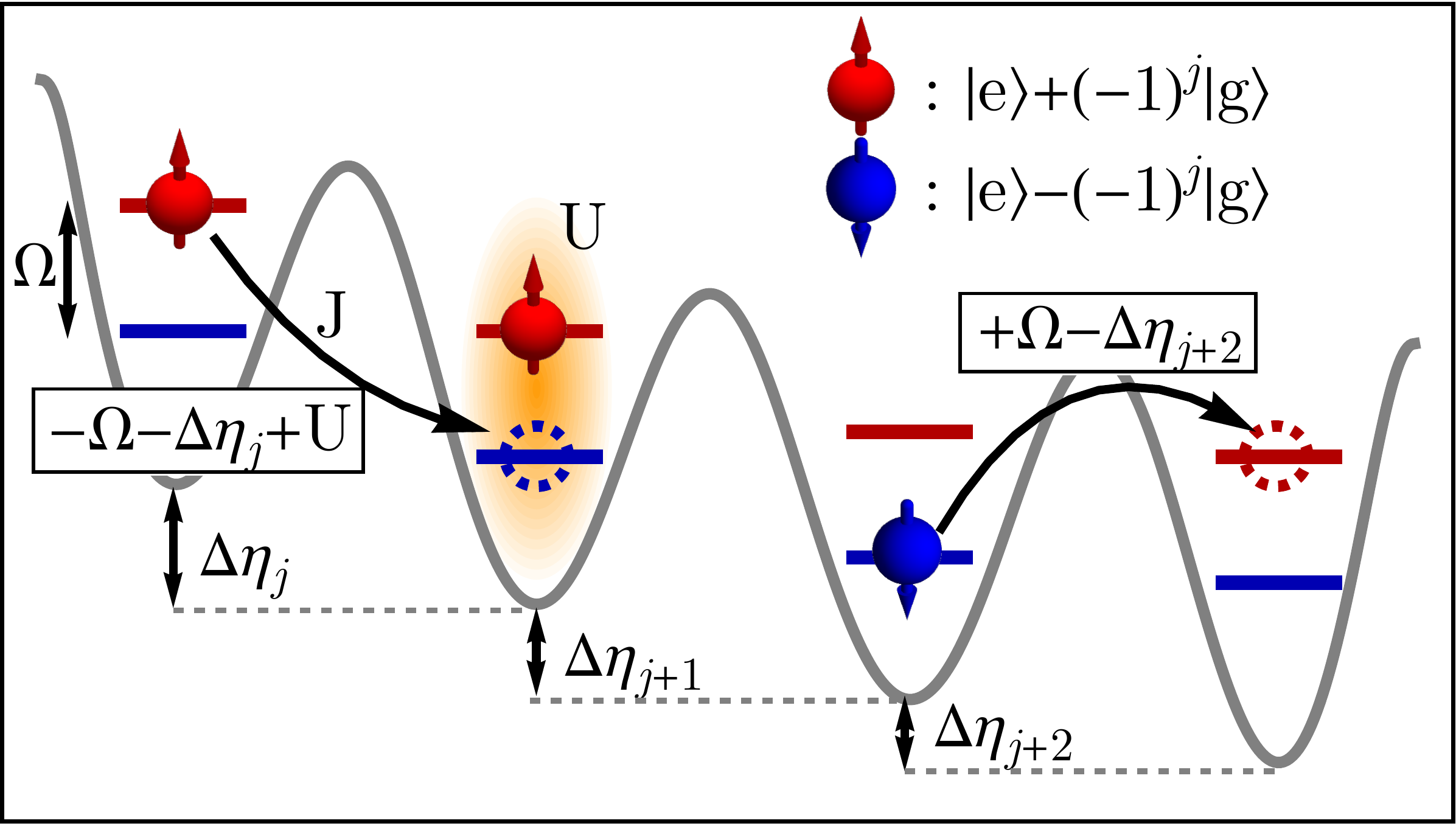}
\caption{Schematic of the optical lattice system, confined to 1D. The red and blue-labeled single-particle eigenstates of the collective drive field are superpositions of bare atomic states $\{g,e\}$, alternating due to the $e^{i j \pi}=(-1)^{j}$ SOC phase in the drive. Atoms tunnel at rate $J$ accompanied with a spin-flip due to the alternating basis. Tunneling incurs energy costs from the trap gradient ($\pm \Delta \eta_j$), atomic interactions (set by $U$) and driving (set by $\Omega$).}
\label{fig_Schematic}
\end{figure}
Despite having site-resolved atomic motion, we do not require site-resolved focused lasers, instead only needing a collective driving laser. We also require no spin-dependent lattice potentials or lattice modulation. The drive, trap and interactions lead to energetic constraints that only allow tunneling between one lattice site pair at a time, while all other sites are effectively decoupled. Our protocol is also robust to slow global phase drifts of the drive, because the system adiabatically follows the drive's single-particle eigenstates throughout the evolution. After state generation, we describe a method to observe the enhanced phase sensitivity without needing many-body measurements such as parity, by instead implementing a reversal of the generation protocol. Finally, we give an augmentation to the protocol that compensates for holes in the loadout. These features together with scalability make our proposal promising for massive entanglement generation and sensitivity improvements in state-of-the-art sensors.

%%%%%
\textit{Model. }   We consider a laser-driven 3D optical lattice populated by fermionic atoms in the lowest motional band, with two internal spin-like states $\sigma \in \{g,e\}$. We assume strong transverse confinement, restricting tunneling to an array of independent 1D chains each of length $L$ and containing $N$ atoms. Each chain operates in the Mott insulating regime with one atom per site ($N=L$). Similar configurations can be generated in tweezer arrays. Fig.~\ref{fig_Schematic} depicts the setup. The Hamiltonian is
\begin{equation}
\label{eq_FermiHubbardLab}
    \hat{H}=\hat{H}_{\mathrm{Hubbard}}+\hat{H}_{\mathrm{Drive}}+\hat{H}_{\mathrm{Trap}},
\end{equation}
where $\hat{H}_{\mathrm{Hubbard}}=-J\sum_{\langle i,j\rangle,\sigma}(\hat{c}_{i,\sigma}^{\dagger}\hat{c}_{j,\sigma}+h.c.)+U\sum_{j}\hat{n}_{j,e}\hat{n}_{j,g}$ is the Fermi-Hubbard Hamiltonian with nearest-neighbour tunneling rate $J$, repulsion $U$, operator $\hat{c}_{j,\sigma}$ annihilating an atom of spin $\sigma$ on site $j$, and $\hat{n}_{j,\sigma}=\hat{c}_{j,\sigma}^{\dagger}\hat{c}_{j,\sigma}$. The laser $\hat{H}_{\mathrm{Drive}}=\frac{\Omega}{2}\sum_{j}(e^{i j \pi}\hat{c}_{j,e}^{\dagger}\hat{c}_{j,g}+h.c.)$ is a collective driving field. The phase $e^{i j \pi}$ is created by a mismatch between the driving and confining laser wavelengths, corresponding to an effective flux $\phi=\pi$ that induces spin-orbit coupling (SOC)~\cite{wall2016SOC}. We also include the trapping potential $\hat{H}_{\mathrm{Trap}}=\eta_{\mathrm{ext}}\sum_{j}(j-j_0)^2(\hat{n}_{j,e}+\hat{n}_{j,g})$ with trap energy $\eta_{\mathrm{ext}}$ from external harmonic confinement (centered on site $j_0$), approximated as quadratic near the center of the lattice, yielding linear potential differences $\Delta \eta_{j}=-2\eta_{\mathrm{ext}}(j-j_0+1/2)$ between neighbouring sites $j$ and $j+1$ [see Supplementary B].

We assume that the drive frequency is much stronger than the tunneling rate, $\Omega \gg J$. Under this condition, the single-particle eigenstates of the system are set by the drive. We rotate into the basis of these eigenstates by defining new fermions $\hat{a}_{j,\uparrow}=(\hat{c}_{j,e}+e^{i j \pi}\hat{c}_{j,g})/\sqrt{2}$, $\hat{a}_{j,\downarrow}=(\hat{c}_{j,e}-e^{i j \pi}\hat{c}_{j,g})/\sqrt{2}$. The Hubbard and drive Hamiltonians become
\begin{equation}
\begin{aligned}
    \hat{H}_{\mathrm{Hubbard}}&=-J\sum_{\langle i,j\rangle}\left(\hat{a}_{i,\uparrow}^{\dagger}\hat{a}_{j,\downarrow}+h.c.\right)+U \sum_{j}\hat{n}_{j,\uparrow}\hat{n}_{j,\downarrow},\\
    \hat{H}_{\mathrm{Drive}}&=\frac{\Omega}{2}\sum_{j}\left(\hat{n}_{j,\uparrow}-\hat{n}_{j,\downarrow}\right),
\end{aligned}
\end{equation}
with $\hat{n}_{j,\tilde{\sigma}}=\hat{a}_{j,\tilde{\sigma}}^{\dagger}\hat{a}_{j,\tilde{\sigma}}$ for drive eigenstates $\tilde{\sigma} \in \{\uparrow,\downarrow\}$. The tunneling is now accompanied by a spin-flip due to the SOC phase. The trapping potential keeps the same form.

While the tunneling couples the drive eigenstates, actual transfer of atoms will depend on the energy differences between states. Some sample tunneling processes are depicted in Fig.~\ref{fig_Schematic}. A spin-$\uparrow$ atom tunneling down the trap gradient incurs an energy change $-\Delta \eta_j$ from the trap, $-\Omega$ from flipping spin, and $+U$ for creating a doublon (two atoms on one site). A spin-$\downarrow$ atom tunneling instead has a change $+\Omega$ from the drive. If the total change is much larger than $J$, tunneling is suppressed. Furthermore, since the trap energy differences $\Delta \eta_j$ vary from site to site, by making the trap strong ($\eta_{\mathrm{ext}} \gg J$) we can tune the drive frequency $\Omega$ to resonantly enable a single tunnel coupling of a chosen spin between two chosen lattice sites while keeping all other tunneling processes offresonant. This allows for site-resolved control of lattice dynamics without needing a focused laser.

%%%%%
\textit{Generation protocol. }
%%%%%
\begin{figure*}
\centering
\includegraphics[width=1\linewidth]{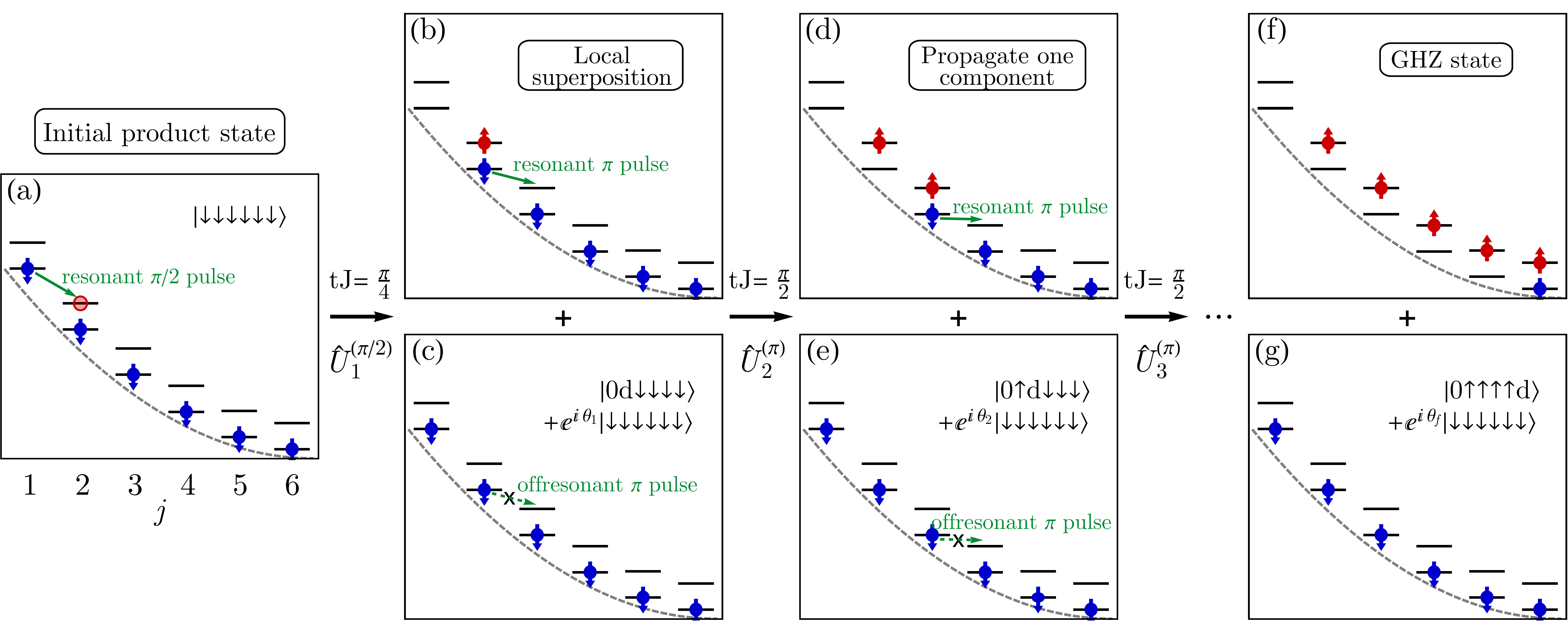}
\caption{Schematic of GHZ state generation protocol. The system is initialized in a product state [panel (a)]. The $j=1$ lattice site has tunneling enabled for its $\downarrow$ atom, and the system is evolved for $tJ=\pi/4$, making an equal-weight superposition [panels (b),(c)]. The panel (b) component is allowed to tunnel further down the lattice, converting $\downarrow$ atoms into $\uparrow$ one site at a time with coherent transfers taking $tJ = \pi/2$ each. The other component [panels (c),(e), corresponding to the initial state] does not evolve because of interaction-induced energy gaps. The end result is a GHZ state [panels (f),(g)]. The insets show the state at each step (with relative phases $\theta_i$, and \textit{d} denoting doublons).}
\label{fig_CatSchematic}
\end{figure*}
The control over tunneling allows us to generate a GHZ state. The scheme is depicted in Fig.~\ref{fig_CatSchematic}. We assume for simplicity that the populated sites do not include the center of the trap potential ($j_0 > L$, with sites indexed $j=1,2,\dots, L$). This can be achieved for example by applying a superimposed linear potential; a trap centered at the middle will be discussed afterwards. We start with a product state $\ket{\psi_0}=\bigotimes_j \ket{\downarrow}_j$ [panel (a)], which can be prepared with a pulse or ramp [see Supplementary A]. The first step is to generate a local two-atom superposition on two adjacent sites, by resonantly enabling the tunneling of the $\downarrow$ atom at site $j=1$ to $j=2$. The drive frequency is set to $\Omega=\Omega_1$, which satisfies $\Omega_1+U-\Delta \eta_1=0$. We keep the laser on with this frequency for a time $tJ = \pi/4$, realizing a unitary operation $\hat{U}_{1}^{(\pi/2)}=e^{-i\hat{H}(\Omega_1) t}$ equivalent to a $\pi/2$ pulse creating an equal-weight superposition of the initial state and a doublon on $j=2$ [panels (b),(c)]. Analogous tunneling from other sites does not occur because other trap energies $\Delta \eta_j$ for $j>1$ differ by at least $2\eta_{\mathrm{ext}} \gg J$.

We next force the $j=2$ site's $\downarrow$ atom to tunnel to $j=3$, but now, set the drive frequency to $\Omega_2$ satisfying $\Omega_2 -\Delta \eta_2 = 0$. The first component of the superposition [panel (b)] will tunnel because it goes from one doublon configuration to another and suffers no penalty $U$. The second component [panel (c)] will have an additional cost $U$, its tunneling will be off-resonant, and it will remain unaltered. We wait a time $tJ = \pi/2$, realizing a unitary $\hat{U}_{2}^{(\pi)}=e^{-i \hat{H}(\Omega_2)t}$ corresponding to a $\pi$ pulse transferring the $\downarrow$ atom from $j=2$ to $j=3$, resulting in a new superposition [panels (d),(e)]. We then make the site $j=3$ doublon have its $\downarrow$ atom tunnel to $j=4$ with another $\pi$ pulse (unitary $\hat{U}_{3}^{(\pi)}$), followed by $j=4$ to $j=5$, repeating to the end of the chain. The final state will take the form,
\begin{eqnarray}
  \ket{\psi_{\mathrm{GHZ}}}&=&\hat{U}_{L-2}^{(\pi)}\dots \hat{U}_{2}^{(\pi)}\hat{U}_{1}^{(\pi/2)}\ket{\psi_0}\\
    &=&\left(\ket{\downarrow,\downarrow,\dots,\downarrow,\downarrow}+e^{i \theta_f}\ket{0,\uparrow,\dots,\uparrow,d}\right)/\sqrt{2},\notag
\end{eqnarray}
as shown in panels (f),(g), corresponding to a GHZ state involving $L$ sites, $L-2$ of which differ in spin projection (still assuming unit filling $N=L$). Here, $\theta_f$ is a relative phase picked up during the evolution [see Supplementary D], and $d$ denotes a doublon. The total evolution time is $tJ = \pi/4 + (L-2)\pi/2$. While the protocol thus far assumed that the chain did not contain the center of the trap, we can also extend it to a symmetric version ($j_0=L/2$). In this case, the superposition will have four components instead of two because each side propagates independently. Such an outcome may be useful in its own right, e.g. to create compass-type states. However, we can also prevent it from happening by disrupting the $\hat{U}_{1}^{(\pi/2)}$ step on one side. Following steps will then fail on that side, allowing the protocol to proceed as before [see Supplementary E].

\begin{figure}
\centering
\includegraphics[width=1\linewidth]{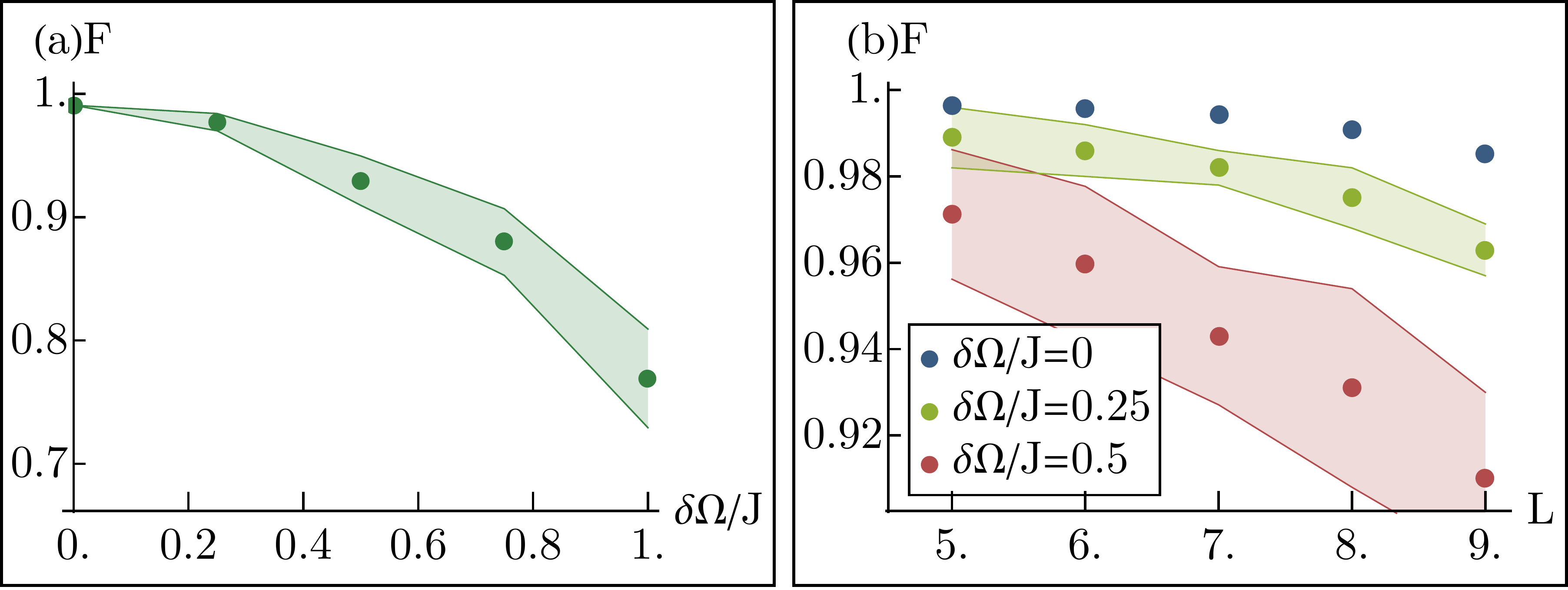}
\caption{(a) Fidelity of the GHZ state with drive frequency noise $\delta \Omega$. The deviation is implemented as a random constant shift $\Omega_i \to \Omega_i+\delta\Omega_i$ to each step of the protocol, uniformly drawn from $\delta\Omega_i\in[-\delta\Omega,\delta\Omega]$ (thus different for every step). Each step is simulated using exact numerical evolution under the full Hamiltonian. Multiple trajectories of randomly-drawn sets $\{\delta \Omega_i \}$ are run, and their fidelity averaged. The shaded region shows one standard deviation. Parameters are $L=8$ at unit filling, $U/J=405$, $\eta_{\mathrm{ext}}/J=21$. (b) Averaged fidelity as a function of $L$, for fixed levels of drive noise.}
\label{fig_Fidelity}
\end{figure}

An important advantage lies in the protocol's piecewise nature. Some methods such as adiabatic dragging suffer from reduced fidelity for larger states due to exponentially shrinking many-body energy gaps with system size. Here, the reduction of the system to an effective two-level configuration at every step allows for easier optimization of the individual steps, and is conceptually straightforward to scale up. Furthermore, the evolving state exhibits some robustness to collective phase-drift effects, e.g. unwanted phases $e^{i \lambda(t)}$ in $\hat{H}_{\mathrm{Drive}}$ for some function $\lambda(t)$. The system will follow the drift by adiabatically remaining in the drive's eigenbasis (provided $\Omega \gg J$ and $\lambda(t)$ varies slowly on the timescale of $\Omega$), preserving the superposition. The main source of error would be imperfect resonance matching $\delta\Omega$ between the desired and actual Rabi frequency $\Omega_i$ at each step. Fig.~\ref{fig_Fidelity} shows a benchmark of the protocol fidelity, averaged over trajectories with random disorder $\delta\Omega$. We see that GHZ states of 10+ sites can be made with fidelities above $90 \%$. Assuming a quadratic decay, we can extrapolate these results to larger states of $L=20$, finding expected fidelities of $F\approx 83 \%$ with $\delta\Omega/J=0.25$ and $F\approx 56 \%$ with $\delta\Omega/J=0.5$. This tolerance can be further improved with a deeper trap, for which the allowed $J$ (and thus mismatch $\delta\Omega$) can be larger.

%%%%%
\textit{Experimental implementation and measurement. }
%%%%%
A feasible platform for our protocol is a 3D optical lattice~\cite{campbell2017clock} or tweezer array~\cite{Norcia2019} loaded with quantum-degenerate fermionic alkaline earth or earth-like atoms such as Sr or Yb. The bare atomic states $\{g,e\}$ can be represented by electronic clock states with optical frequency separation. For a lattice, the confinement should be made strong along transverse directions ($\hat{x}$, $\hat{y}$) and intermediate along the generation direction ($\hat{z}$). A lattice using spin-polarized fermionic $^{87}$Sr at the magic wavelength can realize parameters of $U/J \approx 400$, $\eta_{\mathrm{ext}}/J\approx 20$, $J/(2\pi)\approx 10$ Hz [see Supplementary A]; deeper traps can also be made by reducing beam waist. Note that a deep enough trap can reduce $J$ for sites far from the trap center due to wavefunction deformation, but this reduction should be negligible provided the nearest-neighbour energy differences $\Delta \eta_j$ are much smaller than the band gap. Even if there is a small change to $J$, we only need to run those particular steps for a longer time interval. Typical generation time for these parameters is $t \sim L \times 25$ ms, which is small compared to coherence times $\sim 10$ s~\cite{goban2018coherenceTime} for state size $L \sim 10$ sites. The 3D lattice allows simultaneous creation of many states. from which a constructive measurement signal can be obtained as described below.

\begin{figure}
\centering
\includegraphics[width=1\linewidth]{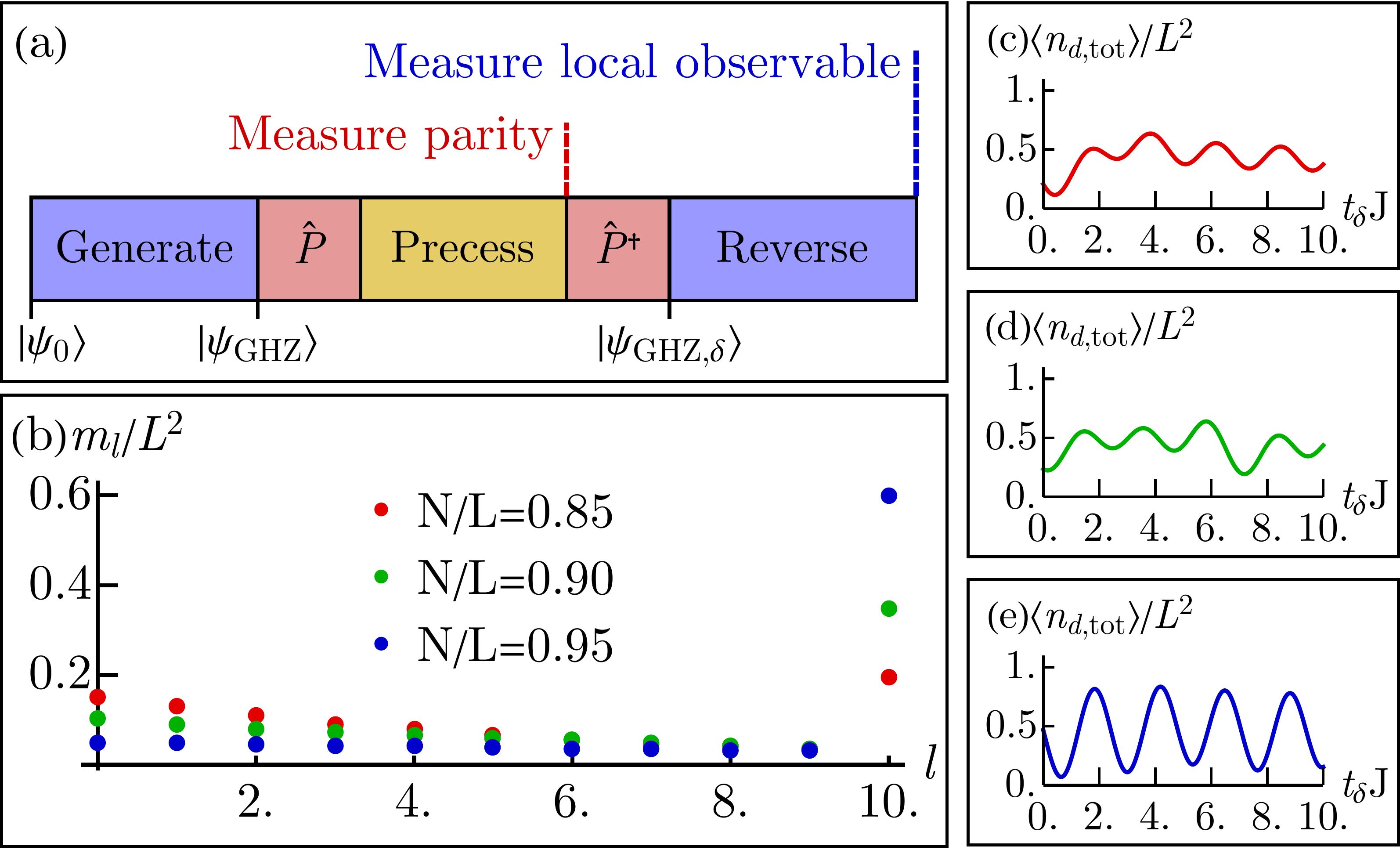}
\caption{(a) Schematic for using the GHZ state in metrology. A parity measurement can be done after allowing the state to precess under detuning $\delta$ (pulse $\hat{P}$ puts the state into the appropriate lab frame [see Supplementary A]). If we reverse the generation after precession, we can instead measure a local observable (doublon number). (b) Average histogram of state lengths that can be generated in a 3D lattice for randomly sprinkled holes given filling fraction $N/L$. Here $m_l$ is the total number of GHZ states of size $l \in [0,1,\dots,L]$ that can be made by starting from one edge of the lattice, counting along a given direction and stopping if we meet a hole. The lattice size is $L \times L \times L$ with $L=10$. (c-e) Sample trajectories of total doublon number [$\hat{n}_{d}$ summed over all states according to the randomly-sprinkled distribution, see Supplementary F] after reversal for different filling fractions $N/L = 0.85,\> 0.90,\> 0.95$ respectively. Detuning is set to $\delta/J = 0.3$. The bare phases $\theta_{r}$ are sampled randomly from $[0,2\pi]$ for simplicity, but made equal for all states of a given length $l$.}
\label{fig_MeasurementDetuning}
\end{figure}

To use the GHZ state for enhanced sensing, we allow it to pick up a relative phase from laser detuning, which is an additional Hamiltonian term $\frac{\delta}{2} \sum_{j}\left(\hat{n}_{j,e}-\hat{n}_{j,g}\right)$. The scheme is depicted in Fig.~\ref{fig_MeasurementDetuning}(a). After generation, a pulse $\hat{P}$ rotates the state into a form where its superposition components will acquire a relative phase $\theta_{\delta}=\delta (N-1) t_{\delta}$ if they precess for time $t_{\delta}$ [$N-1$ because of the edge sites, see Supplementary F]. Conventionally, this $N$-proportional enhancement is observed using a Ramsey sequence followed by a parity measurement~\cite{boyd2006ramsey,ramsey2005ramsey,huang2015quantumMetrology}, requiring measurement of $N$-body correlators which can be challenging for clocks, although it can be done in tweezers~\cite{Norcia2019}.

As an alternative approach, we instead undo the generation sequence, as shown in Fig.~\ref{fig_MeasurementDetuning}(a). After precession, we rotate the state back into the gauged frame with another pulse $\hat{P}^{\dagger}$ [see Supplementary F]. We then do the $\pi$-pulse steps in reverse order, $\hat{U}_{2}^{(\pi)}\dots\hat{U}_{L-2}^{(\pi)}\ket{\psi_{\mathrm{GHZ,\delta}}}$ (with $\ket{\psi_{\mathrm{GHZ,\delta}}}$ the state after precession and applying $\hat{P}^{\dagger}$). These steps reduce the state to $(\ket{\downarrow,\downarrow}+e^{i(\theta_{r}+\theta_{\delta})}\ket{0,d})/\sqrt{2}\>\otimes \ket{\downarrow,\dots,\downarrow}$, where the superposition is back on two sites $j=1,2$ and $\theta_{r}$ is a constant phase depending on system size and parameters. Reapplying unitary $\hat{U}_{1}^{(\pi/2)}$ will rotate this state into a form where the relative phase may be measured from doublon number $\langle \hat{n}_{d}\rangle = \sum_{j}\langle\hat{n}_{j,\uparrow}\hat{n}_{j,\downarrow}\rangle$ in the vicinity of $j=1,2$, without needing $N$-body correlators. The doublon number will oscillate as a function of $t_{\delta}$, allowing the detuning to be obtained from the period.

We have assumed unit filling. While unwanted holes in a 3D lattice will be confined by the energy gaps, they will interrupt state generation, leading to GHZ states of different sizes. However, sufficiently high filling will allow the maximum-length ones to dominate the signal. We benchmark the measurement protocol in Fig.~\ref{fig_MeasurementDetuning}(b-e) by randomly sprinkling holes into a 3D lattice, and computing how many states of each length we get. Panel (b) shows the distribution of number $m_l$ for state size $l \in [0,1,\dots L]$ while panels (c-e) give sample oscillation trajectories of total doublon number $\langle \hat{n}_{d,\mathrm{tot}}\rangle$ [$\langle \hat{n}_{d}\rangle$ summed over the array of states, SOM]. For $L=10$, fillings above $N/L \gtrsim 0.9$ yield a clear oscillatory signal $(10-1)$ times faster than a single unentangled atom, leading to $\sqrt{10-1}$ times faster clock protocols~\cite{bollinger1996optimal}. One may also employ Fourier analysis to discern the contributions of different sizes.

%%%%%
\textit{Hole correction protocol. }
%%%%%
Our protocol can be modified to compensate for small numbers of holes at the cost of longer generation time. We can augment every step of the original protocol except the first with two auxiliary steps. We first attempt to move a $\downarrow$ atom to make a doublon on the next lattice site as normal ($\ket{d,\downarrow} \to \ket{\uparrow,d}$). If an atom is missing, $\ket{d,0}$, this step will fail. We then apply an auxiliary step that repeats the same tunneling process, but now assuming the target site to have no atom, allowing the transfer $\ket{d,0} \to \ket{\uparrow,\uparrow}$. A second auxiliary step moves the remaining atom over, $\ket{\uparrow,\uparrow}\to \ket{0,d}$, and the protocol may continue. If no holes were present, neither auxiliary step would have an effect because they would be off-resonant [see Supplementary C]. Note that the all-spin-$\downarrow$ superposition component will also suffer local changes, but these will not propagate further, maintaining a significant difference in spin projection [see Supplementary C]. While this augmentation is not as useful to 3D lattice setups whose measurement signal comes from the largest-size state, it can be useful for tweezer systems.

%%%%%
\textit{Conclusions. }
%%%%%
We have proposed a method for generating GHZ states with ultracold fermions that can be directly implemented with 3D lattice systems or tweezer arrays. The resulting state can be immediately used \textit{in situ} for metrological purposes through a Ramsey-like sequence combined with protocol reversal. The fidelity requires good control over drive frequency, but this requirement can be made less stringent with a stronger trap, which also allows for larger tunneling rates and faster generation time. With a 2D tweezer array, one could even generate a GHZ state along one 1D tube, then repeat the protocol along a transverse axis, leading to a 2D GHZ state. One may also use a purification scheme to convert many bad GHZ states into a smaller number of good ones~\cite{dur2003purification}. Altogether, this scheme offers a promising way to generate and use strongly entangled states in metrologically relevant systems.

%%%%%
\textit{Acknowledgements. }
%%%%%
M.M. acknowledges a CTQM graduate fellowship. This work is supported by the AFOSR grant FA9550-19-1-0275, by the DARPA and the ARO grant W911NF-16-1-0576, the NSF grant PHY1820885, NSF JILA-PFC PHY-1734006 grants, and by NIST.

\bibliography{CatBibliography}
\bibliographystyle{unsrt}

\clearpage

\onecolumngrid
\renewcommand{\thesubsection}{\Alph{subsection}}
\renewcommand{\appendixtocname}{Supplementary material}
\appendixpageoff
\appendixtitleoff
\begin{appendices}

\begin{center}
\LARGE
Supplementary Material
\normalsize
\end{center}

%%%%%%%%%%%%%%%%%%%%%%%
\section{State rotations and preparation}
\renewcommand{\thefigure}{A\arabic{figure}}
\renewcommand{\theequation}{A\arabic{equation}}
\setcounter{figure}{0}
\setcounter{equation}{0}
%%%%%%%%%%%%%%%%%%%%%%%
The GHZ state generating protocol requires an initial state of $\ket{\psi_0}=\bigotimes_{j}\ket{\downarrow}_{j}$ in the basis of the drive eigenstates. Such a state has a nontrivial spin structure in the bare atomic state basis $\{g,e\}$ due to the alternating sign of the drive; if we enumerate the lattice sites as $j=1,2,\dots$, the state would be written in the lab frame as,
\begin{equation}
    \ket{\psi_0}=\ket{+x,-x,+x,-x,\dots},
\end{equation}
where $\ket{\pm x} = (\ket{e}\pm \ket{g})/\sqrt{2}$. Preparing this state can be done in two ways. The first is to use a pulse from a laser with the same SOC spatially-varying phase $e^{i j \pi}$, but with an overall phase shift from the drive laser used in the main protocol. Recalling that the drive laser Hamiltonian is,
\begin{equation}
    \hat{H}_{\mathrm{Drive}}= \frac{\Omega}{2}\sum_{j}\left(e^{i j \pi}\hat{c}_{j,e}^{\dagger}\hat{c}_{j,g}+h.c.\right),
\end{equation}
the pulse laser would need to be of the form,
\begin{equation}
    \hat{H}_{\mathrm{P}}= \frac{\Omega_{\mathrm{P}}}{2}\sum_{j}\left(e^{i j \pi - i \pi/2}\hat{c}_{j,e}^{\dagger}\hat{c}_{j,g}+h.c.\right).
\end{equation}
Note that the drive laser's overall phase besides the SOC does not matter as the system will follow the drive's eigenstates; the pulse laser only needs to have its phase behind that of the drive laser by $\pi/2$. One can do both the pulse and driving with the same laser setup since only one beam needs to be active at a time; a mirror and switching configuration can first enable the pulse, followed by the drive for the main steps of the protocol.

The initial state can be prepared by first loading the atoms into their natural ground-state $\bigotimes_{j}\ket{g}_{j}$ in the lab frame by standard cooling techniques, then making a fast pulse,
\begin{equation}
\begin{aligned}
\hat{P}&=e^{-i \frac{\pi}{2\Omega_{\mathrm{P}}}\hat{H}_{\mathrm{P}}},\\
\ket{\psi_0}&=\hat{P}\left(\bigotimes_{j}\ket{g}_{j}\right).
\end{aligned}
\end{equation}
assuming that $\Omega_{\mathrm{P}}\gg J$ to avoid unwanted lattice dynamics. Once this is done the pulse laser is turned off, the drive laser enabled, and the generation protocol may proceed. The same pulse $\hat{P}$ may be used to rotate the final GHZ state into a form where its components will accrue opposite phases from any laser detuning, as described in the measurement protocol of the main text:
\begin{equation}
    \hat{P}\ket{\psi_{\mathrm{GHZ}}}=(\ket{g,g,\dots,g,g}+e^{-i\theta_f}\ket{0,e,\dots,e,d})/\sqrt{2}.
\end{equation}

\begin{figure}
\centering
\includegraphics[width=0.4\linewidth]{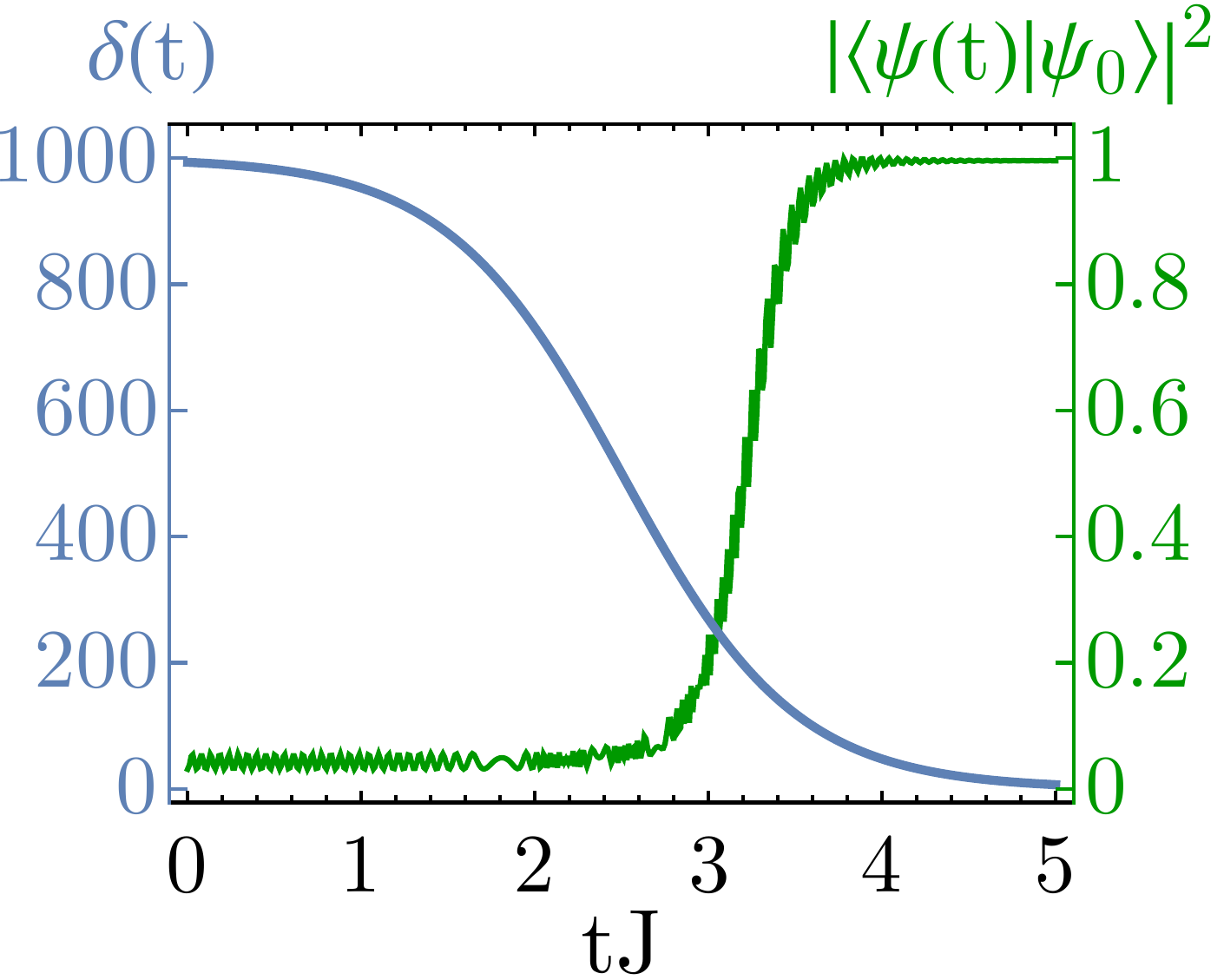}
\caption{Initial-state $\ket{\psi_0}$ preparation using an adiabatic ramp of the detuning. Parameters are $t_{\mathrm{ramp}}J=5$, $\delta_0/J=1000$, $\Omega/J=122$, $U/J=405$, $\eta_{\mathrm{ext}}/J=21$, and $L=5$ (trap centered on $j_0 = 3$). The drive frequency $\Omega$ is purposefully chosen to avoid any resonances, preventing unwanted tunneling during this step.}
\label{fig_DetuningPreparation}
\end{figure}

An alternative method for preparing the initial state is to instead use an adiabatic ramp. For this, we only use the drive laser with no need for a pulse. Recall that the drive may have a detuning,
\begin{equation}
    \hat{H}_{\delta}=\frac{\delta}{2}\sum_{j}\left(\hat{n}_{j,e}-\hat{n}_{j,g}\right).
\end{equation}
If the detuning is much larger than the drive frequency, $\delta \gg \Omega$, then the ground-state of the system will be $\bigotimes_{j}\ket{g}_{j}$ even in the presence of the drive. We slowly reduce the detuning from $\delta_0 \gg \Omega$ to zero over a time $t_{\mathrm{ramp}}$, as depicted in Fig.~\ref{fig_DetuningPreparation}:
\begin{equation}
\delta(t) = \delta_{0}\left[\tanh \big(\left(\frac{t_{\mathrm{ramp}}}{2}-t\right)J\big)-1\right].
\end{equation}
The system will adiabatically remain in the ground-state, which will transition from $\bigotimes_{j}\ket{g}_{j}$ to $\ket{\psi_0}$, provided that the rate $d\delta/dt$ is smaller than the gap to the next-lowest energy state proportional to $\Omega$, and $\Omega$ is chosen to avoid any tunneling resonances.

%%%%%%%%%%%%%%%%%%%%%%%
\section{System parameters}
\renewcommand{\thefigure}{B\arabic{figure}}
\renewcommand{\theequation}{B\arabic{equation}}
\setcounter{figure}{0}
\setcounter{equation}{0}
%%%%%%%%%%%%%%%%%%%%%%%
In this section we give an overview of the sample parameters used throughout the main text, assuming realistic experimental setups. We consider a 3D optical lattice loaded with nuclear-spin polarized fermionic $^{87}$Sr at the magic wavelength $\lambda_{\mathrm{L}}\approx 813$ nm (lattice constant $a=\lambda_{\mathrm{L}}/2)$, with the clock states $^{1}S_{0}$, $^{3}P_{0}$ acting as the bare spin states $g,e$. The 3D lattice confinement strengths are set to $(V_x,V_y,V_z)=(200,200,19)E_r$, with $E_r=\pi^2\hbar^2/(2m a^2)$ the recoil energy ($m\approx 87$ amu). The desired parameter regimes are $U, \eta_{\mathrm{ext}} \gg J$.

\begin{figure}
\centering
\includegraphics[width=0.4\linewidth]{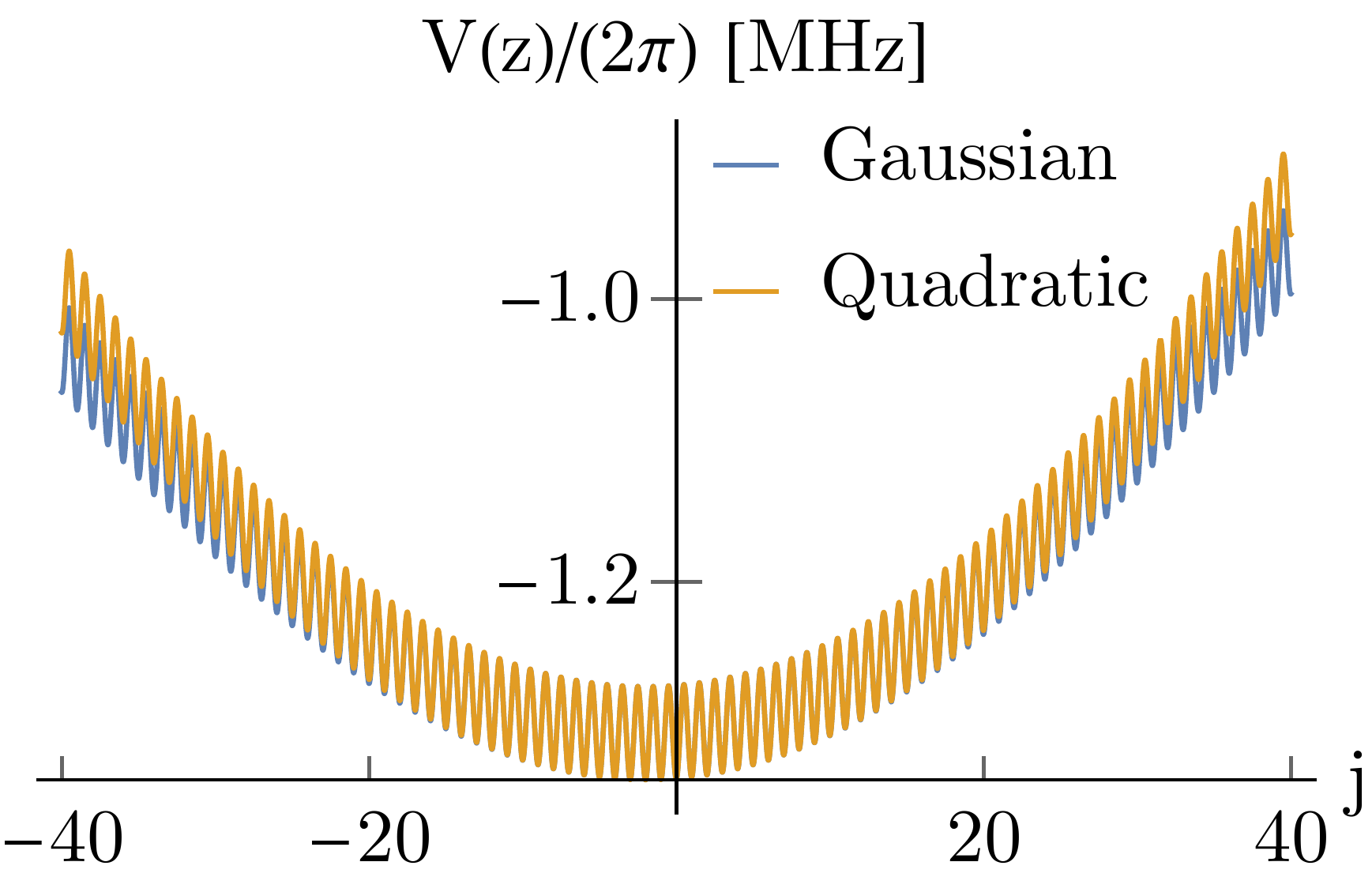}
\caption{Schematic of lattice trapping potential along the $\hat{z}$ direction which we use to make GHZ states, in units of site number $j=z/a$. The full Gaussian profile [Eq.~\eqref{eq_PotentialZ}] and its approximate quadratic form [Eq.~\eqref{eq_PotentialZQuadratic}] are shown. The center is shifted by gravity, but only by a few sites $j_0 \approx -2$. We see that the quadratic approximation remains valid for $\sim 40$ sites.}
\label{fig_Trap}
\end{figure}

The GHZ state is generated along the $\hat{z}$ direction. While there will be a gravitational potential shift, its only effect will be to move the center of the trap by a few lattice sites. This does not affect our protocol provided that we account for the shift when determining the resonant drive frequency for the first step $\hat{U}_{1}^{(\pi/2)}$ (other steps only deal with relative energies that are insensitive to the center of the trap). Assuming Gaussian lattice beam waists of $\nu_{x}= \nu_{y} = 45$ $\mu$m along $\hat{z}$ from the transverse $\hat{x}$, $\hat{y}$ directions, the potential along $\hat{z}$ will be given by,
\begin{equation}
    V(z)=V_{z}\sin^2\left(\frac{\pi z}{a}\right)+m g z-\tilde{V}_{x}e^{-\frac{2z^2}{\nu_x}}-\tilde{V}_{y}e^{-\frac{2z^2}{\nu_y}},
\end{equation}
where $\tilde{V}_{x}=V_{x}-\sqrt{V_x E_r}/2$, $\tilde{V}_{y}=V_{y}-\sqrt{V_{y} E_r}/2$ are renormalized lattice depths, and $g$ is gravitational acceleration. Fig.~\ref{fig_Trap} shows this potential as a function of lattice site number $j$ (i.e. in units of $z/a$). The Gaussian profile can be approximated by a quadratic function near the bottom,
\begin{equation}
\label{eq_PotentialZ}
    V(z)\approx V_{z}\sin^2\left(\frac{\pi z}{a}\right)+m g z-(\tilde{V}_{x}+\tilde{V}_{y})+\left(\frac{2\tilde{V}_x}{\nu_x^2}+\frac{2\tilde{V}_y}{\nu_y^2}\right)z^2.
\end{equation}
As seen from Fig.~\ref{fig_Trap}, this approximation works well for $\sim 40$ sites nearest to the center of the trap. The first term creates the lattice potential built into our Fermi-Hubbard model. The last term's prefactor sets the trap energy (normalizing by the lattice constant),
\begin{equation}
\label{eq_PotentialZQuadratic}
    \eta_{\mathrm{ext}}/(2\pi)=\frac{2V_x}{(\nu_x/a)^2}+\frac{2V_y}{(\nu_y/a)^2} \approx 219 \text{ Hz.}
\end{equation}
The gravitational potential $mgz$ creates a shift, $j_0 = -\eta_{\mathrm{ext}}/(2mga) \approx -2$ sites, which may be accounted for when choosing the drive frequency for step $\hat{U}_{1}^{(\pi/2)}$.

We also evaluate the tunneling overlap integral and onsite s-wave interaction strength (for scattering length $a_{eg}^{-} = 69.1 a_0$ with $a_0$ the Bohr radius) via standard Wannier orbital calculations, yielding,
\begin{equation}
    J/(2\pi) \approx 10.4 \text{ Hz},\>\> U/(2\pi) \approx 4212 \text{ Hz}.
\end{equation}
From the above, we conclude that our system parameters in units of $J$ are given by,
\begin{equation}
    U/J\approx 405, \>\>\> \eta_{\mathrm{ext}}/J\approx 21.
\end{equation}
The drive Rabi frequency can be made on the order of kHz, yielding possible values $\Omega/J \sim 1-1000$.

Note that the above approximations are limited by our assumption of perfect spatial Wannier orbitals. Lattice sites far away from the trap center can have their spatial wavefunctions deformed and tunneling rate $J$ reduced if the local trap energy differences $\Delta \eta_j =-2 \eta_{\mathrm{ext}}(j-j_0+\frac{1}{2})$ approach the band gap (also leading to unwanted band-changing tunneling). For our parameters, we can still be $\sim 10$ sites from the center (thus a 20-site cat) and have $|\Delta \eta_j| \approx 4$ kHz compared to the band gap $\omega_z/(2\pi\hbar) \approx 26$ kHz, for which ground-band tunneling $J$ is changed by less than 1\% and band-changing tunneling rates are $< 0.1$ Hz (very offresonant due to the energy difference of $\approx 22$ kHz). Even if there was a notable reduction of $J$, the protocol can still function - the individual steps will just need to take a longer amount of time to compensate.

As a side note, in the above parameters we have also ensured that $U \gg \eta_{\mathrm{ext}}$. This is not strictly necessary, and is done to ensure that no accidental resonances occur with lattice sites not involved in the current active step of the protocol (many such unwanted resonances are shifted by $U$, and can thus be enabled by accident if $U \approx \Delta \eta_{j}$ for some $j$ uninvolved in the current step). For larger $L$ where the trap energy differences $\Delta \eta_{j}$ grow large, one can instead dodge unwanted resonances by tuning $U$ between them. It is straightforward to analytically compute all possible resonant drive frequencies for all tunneling events at every step, and determine experimentally-appropriate values of $U$, $\eta_{\mathrm{ext}}$ for which the drive frequency can isolate the desired resonance while being sufficiently far from all others.

%%%%%%%%%%%%%%%%%%%%%%%
\section{Hole correction protocol}
\renewcommand{\thefigure}{C\arabic{figure}}
\renewcommand{\theequation}{C\arabic{equation}}
\setcounter{figure}{0}
\setcounter{equation}{0}
%%%%%%%%%%%%%%%%%%%%%%%

\begin{figure}
\centering
\includegraphics[width=0.6\linewidth]{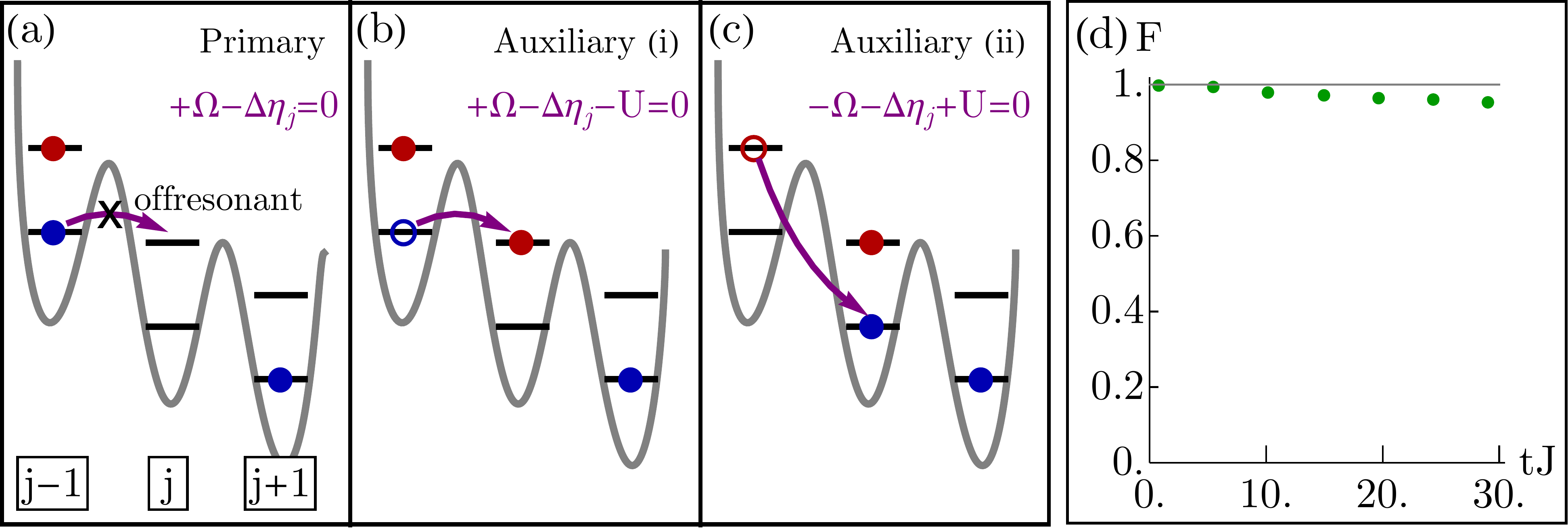}
\caption{Schematic for the hole-correcting protocol. Panel (a) shows the primary step of the regular protocol, which will fail if site $j$ has a hole. Panels (b),(c) give two auxiliary steps (i) and (ii), which compensate for the hole by manually moving the doublon on site $j-1$ into the hole's location so that the protocol can keep going. If the primary step in panel (a) had succeeded, the auxiliary steps would have no effect. Panel (d) shows the fidelity of obtaining the desired state after every primary-auxiliary-auxiliary sequence of the protocol, using a numerical evolution of size $L=8$ with $N=7$ and a single hole at $j=4$. The other component of the superposition will see local changes near the hole, but will otherwise be unaffected.}
\label{fig_Error}
\end{figure}

In this section we provide details for the hole correction protocol described in the main text. Fig.~\ref{fig_Error}(a-c) shows a schematic diagram. Every primary step  [panel (a), moving a doublon over one site by making its $\downarrow$ atom tunnel] is followed by two auxiliary steps [panels (b),(c)], whose combined effect is to manually move the doublon over if there was a hole present and the primary step failed. Fig.~\ref{fig_Error}(d) shows fidelities of generating the desired state after every primary-auxiliary-auxiliary sequence for a sample GHZ state. Note that the all-spin-$\downarrow$ superposition component also suffers local changes in the vicinity of the hole as described below; the fidelities quoted account for this by assuming the superposition to consist of the two Fock states that the resonant tunneling processes are expected to create after each primary-auxiliary-auxiliary sequence. Note also that we do not account for holes on the first two sites $j=1,2$ used to build the initial superposition, although an analogous sequence could be designed for that step as well.

\begin{figure}
\centering
\includegraphics[width=0.55\linewidth]{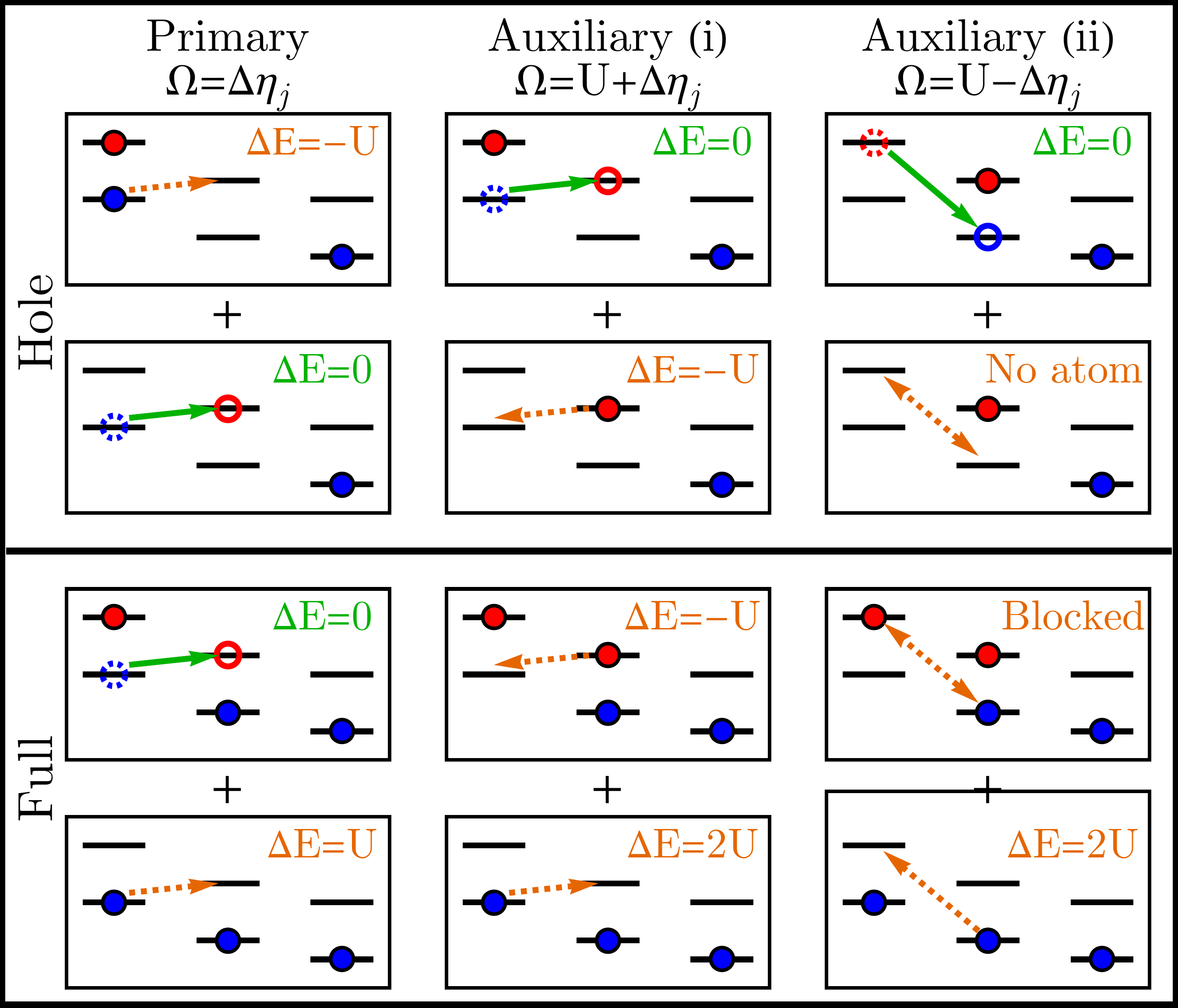}
\caption{Schematic for the hole correction protocol. The primary step of the original protocol is followed by two auxiliary steps (i), (ii). The top panel depicts the superposition that would exist if a vacancy was present (Hole), while the bottom panel shows unit filling (Full). For every step, a green arrow means the drive is resonant with the tunneling process in question, and an atom is moved over with a $\pi$ pulse. For an orange dashed arrow, the process is either offresonant or otherwise inhibited. For unit filling, only the primary step succeeds, moving a doublon over one site. For the hole, the primary step fails, but the two auxiliary steps move the doublon over manually so the protocol may continue. Note that in the case of the hole, there is also a spin-flip on the all-$\downarrow$ component which we do not want to affect, reducing the relative difference in spin projection. However, this change is not propagated further.}
\label{fig_HoleCorrection}
\end{figure}

Fig.~\ref{fig_HoleCorrection} depicts the protocol in more detail. We compare the situation where a hole is present on the site we want to move the doublon into (top half), with the situation of unit filling where only the primary step should take effect (bottom half). For each step, the drive frequency is shown, as well as the two components of the superposition. Steps where the atom tunneling will succeed are shown in green; for those, the Rabi frequency satisfies the respective resonance condition, and the total energy difference before/after tunneling is $\Delta E = 0$. Steps where tunneling fails are shown in orange; for these, either $|\Delta E| \sim U \gg J$, there are no atoms in the coupled levels, or atoms populate both levels and are Pauli blocked.

The price we pay aside from increased evolution time is that the other (no-doublon) component of the superposition will now have an additional $\uparrow$ atom, whereas we want it to be all $\downarrow$. However, this will not propagate further, and assuming a small density of holes we should still have mostly $\downarrow$ atoms in the no-doublon component, and mostly $\uparrow$ atoms in the evolving one.

%%%%%%%%%%%%%%%%%%%%%%%
\section{Relative phase from unperturbed generation}
\renewcommand{\thefigure}{D\arabic{figure}}
\renewcommand{\theequation}{D\arabic{equation}}
\setcounter{figure}{0}
\setcounter{equation}{0}
%%%%%%%%%%%%%%%%%%%%%%%
In this section, we give the relative phase that the GHZ state picks up during the generation protocol. The first step only yields a phase of $e^{i \pi/2}$ between the superposition components due to the $\pi/2$ pulse, because the relative energies of the coupled states are manually set to be equal by choice of drive frequency. For all subsequent steps, while the energies of the two sites tunneling are still matched, the superposition components will have other uninvolved lattice sites with different spin structure (as part of our entanglement-building process), thus picking up a phase from the drive at different rates. This relative phase will vary from step to step, because both the Rabi frequency and the number of misaligned spins between the two superposition components will change.

For a system of $L$ sites and $N=L$ atoms, we have 1 superposition-generating $\pi/2$ pulse, followed by $L-2$ atom-transferring $\pi$ pulses. The total relative phase for the GHZ state,
\begin{equation}
    \ket{\psi_{\mathrm{GHZ}}}=\left(\ket{\downarrow,\downarrow,\dots,\downarrow,\downarrow}+e^{i\theta_f}\ket{0,\uparrow,\dots,\uparrow,d}\right)/\sqrt{2},
\end{equation}
may be found after some algebra to be,
\begin{equation}
\theta_f = \frac{\pi}{2}\left[(L-1)-\frac{U}{J}(L-2)+\frac{\eta_{\mathrm{ext}}}{J}\frac{L}{3}(L-1)(L-2)\right].
\end{equation}
Any measurement protocol would create an additional shift to this overall phase. Note that this result is only exact in the limit where the energy gaps to all unwanted resonances are infinite. For realistic experimental parameters, there may be some deviation to the above with larger states. However, for a measurement protocol such as the reversal described in the main text, this relative phase is unimportant anyways; we only provide it for completeness.

%%%%%%%%%%%%%%%%%%%%%%%
\section{Two-sided trap}
\renewcommand{\thefigure}{E\arabic{figure}}
\renewcommand{\theequation}{E\arabic{equation}}
\setcounter{figure}{0}
\setcounter{equation}{0}
%%%%%%%%%%%%%%%%%%%%%%%
\begin{figure}
\centering
\includegraphics[width=0.35\linewidth]{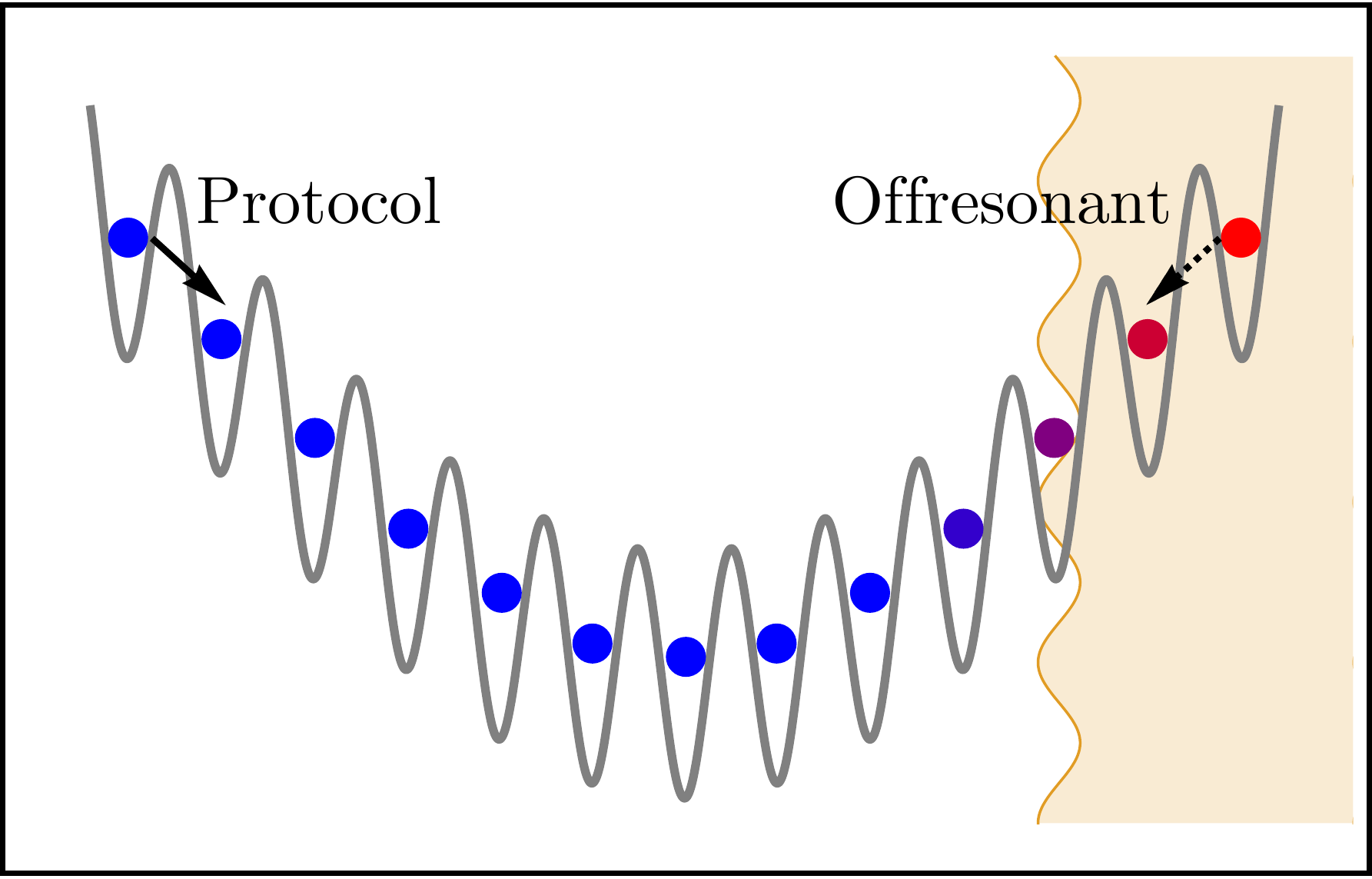}
\caption{Schematic of two-sided trap implementation. A pulse laser (not necessarily site-resolved) can disrupt the protocol on the right side by flipping the spin, preventing the first step from succeeding there. This allows implementation of the protocol from the left with only two superposition components. In principle one can continue through the center and towards the right, using any unaffected (still $\downarrow$) sites there.}
\label{fig_TwoSided}
\end{figure}
Our generation protocol can be generalized to include both halves of the harmonic trap. With the right half included the first step will generate a four-component superposition instead of two, because assuming the center of the trap $j_0$ is an integer, the left and right sides will have identical resonant tunneling and generate independent two-component superpositions (a four-component tensor product overall). The following steps will propagate these superpositions down the lattice on their respective sides independently, at least until we reach the very bottom. We can prevent this from happening by modifying the state preparation. One simple way is to shift the trap potential so that its center is closer to the edge of the atomic cloud (i.e. $|j_0| \gg 1$). Another way is to use a narrow beam-waist laser to effect a $\pi$ pulse on the atoms in the upper-right half of the trap after preparing the $\downarrow$ product state, shown in Fig.~\ref{fig_TwoSided}. This does not need to be single-site focused or fully coherent; we simply need to disrupt the state of the right-side lattice site at the height of the left-side starting point, so that it cannot participate in the protocol's first step. Collateral changes to neighbouring sites on the right are also acceptable, so long as they do not stretch across the whole lattice. With this done, the protocol will fail to start on the right side. Further steps will also fail as they are contingent upon one another. We can then enact the protocol from the left side as before. In principle, we can even continue through the center and out to any unchanged sites on the right.

%%%%%%%%%%%%%%%%%%%%%%%
\section{GHZ state measurement through unitary reversal}
\renewcommand{\thefigure}{F\arabic{figure}}
\renewcommand{\theequation}{F\arabic{equation}}
\setcounter{figure}{0}
\setcounter{equation}{0}
%%%%%%%%%%%%%%%%%%%%%%%

In this section we detail the way to measure the relative phase between the components of the GHZ state through time-reversal. We assume that the state is generated, and allow it to accrue a phase during precession time $t_{\delta}$ from detuning. The drive frequency $\Omega$ is either turned off or tuned to some value far from any resonances during this time, to help prevent the atoms from tunneling. We use the pulse described in Section A to put the GHZ state into the lab frame before the precession starts, and convert it back into the drive frame after the precession, so that its components can accrue the maximum possible phase. This precession may be written as,
\begin{equation}
    \ket{\psi_{\mathrm{GHZ},\delta}}=\hat{P}^{\dagger}e^{-i(\hat{H}+\hat{H}_{\delta})t_{\delta}}\hat{P}\ket{\psi_{\mathrm{GHZ}}},
\end{equation}
which will yield a state of the form,
\begin{equation}
   \ket{\psi_{\mathrm{GHZ},\delta}}=(\ket{\downarrow,\downarrow,\dots,\downarrow,\downarrow}+e^{i(\theta_{p}+\theta_{\delta})}\ket{0,\uparrow,\dots,\uparrow,d})/\sqrt{2},
\end{equation}
where $\theta_{p}$ is a bare phase coming from precession under the drive and interactions, and $\theta_{\delta}=\delta(L-1) t_{\delta}$ is the additional phase from the detuning (minus one because of the edge sites). Note that we have not provided an explicit expression for $\theta_{p}$, which will depend on the system parameters and precession time. However if we emulate a Ramsey-type sequence and the filling fraction is sufficiently high, this bare phase will be irrelevant so long as it is the same for all 1D chains, because we only care about the period of resulting oscillations.

We now run the generating protocol on $\ket{\psi_{\mathrm{GHZ},\delta}}$ in reverse. All of the $\pi$-pulse transfer steps (i.e. all except the first step) are done the same way as the original protocol, only in opposite order. After doing all the steps except $\hat{U}_{1}^{(\pi/2)}$, the result will be,
\begin{equation}
\begin{aligned}
\ket{\psi_{\mathrm{GHZ,r}}}&=\hat{U}_{2}^{(\pi)}\cdots \hat{U}_{L-2}^{(\pi)}\ket{\psi_{\mathrm{GHZ},\delta}}\\
&=\frac{1}{\sqrt{2}}\left[\ket{\downarrow,\downarrow}+e^{i(\theta_{r}+\theta_{\delta})}\ket{0,d}\right]_{j=1,2}\otimes \ket{\downarrow,\dots,\downarrow}_{j=3\dots L}.
\end{aligned}
\end{equation}
Again, the bare phase $\theta_r$ after reversing will be nontrivial even for no precession $t_{\delta}=0$ because applying the steps in reverse does not constitute a true many-body unitary reversal. However, for a given state length and set of parameters it should be the same for every experiment shot.

At this point, we have reduced the system back into a two-state configuration where the relative phase can be measured through a direct laser coupling. The final step is to reapply $\hat{U}_{1}^{(\pi/2)}$,
\begin{equation}
\hat{U}_{1}^{(\pi/2)}\ket{\psi_{\mathrm{GHZ},r}}=\frac{1}{2}\left[(1-i e^{i(\theta_r+\theta_{\delta})})\ket{\downarrow,\downarrow}-i(1+ie^{i(\theta_r+\theta_{\delta})})\ket{0,d}\right]_{j=1,2}\otimes \ket{\downarrow,\dots,\downarrow}_{j=3\dots L},
\end{equation}
for which the relative phase can be obtained from the overlap with the second state of the superposition, equivalent to measuring the doublon number. We only count doublons in the vicinity of the initial site where the original protocol was started, but do not need single-site resolution; a few sites' width is fine, as if the protocol started at $j=1$, we expect the doublon at $j=2$. Provided any unwanted resonances are avoided and the lattice is sufficiently deep, the only way a doublon could be created in this vicinity is if the generating protocol reversed itself as described. Measuring the doublon number for this state yields,
\begin{equation}
\label{eq_AppDoublonNumber}
    \langle \hat{n}_{d}\rangle = \sum_{j}\langle \hat{n}_{j,\uparrow}\hat{n}_{j,\downarrow}\rangle= \frac{1}{2}[1-\sin(\theta_{r}+\theta_{\delta})],
\end{equation}
which contains the $L$-proportional phase accrued from detuning.

Of course, when holes are present, not all GHZ states will be of the desired length $L$. To estimate the signal, we randomly sample a 3D lattice of dimensions $L\times L \times L$ by sprinkling holes to a desired filling fraction $N/L$. We then compute a histogram of the distribution of the number of states $m_{l}$ with length $l \in [0,1,\dots, L]$ that one can realize with this lattice along a given direction $\hat{z}$. For example, if a given 1D tube has a hole at the $8$th site from the bottom, that tube adds one to $m_{7}$. Main text Fig.~\ref{fig_MeasurementDetuning}(b) plots this distribution for different filling fractions, finding that for $N/L \gtrsim 0.9$ the majority of the tubes should yield full-length states. We then sum the doublon number from Eq.~\eqref{eq_AppDoublonNumber},
\begin{equation}
\label{eq_AppTotalDoublonNumber}
\langle \hat{n}_{d,\mathrm{tot}}\rangle = \sum_{l=0}^{L}m_{l}\times \frac{1}{2}[1-\sin(\theta_{r}^{(l)}+\delta (l-1) t)].
\end{equation}
The bare phase $\theta_{r}^{(l)}$ is equal for all states of a given size $l$ for fixed system parameters, which allows equal-length states to contribute to the signal constructively. For simplicity, we select a random $\theta_{r}^{(l)}$ for every $l$. The total resulting signal is plotted as a function of $t$ in main text Fig.~\ref{fig_MeasurementDetuning}(b-d). For sufficiently high filling the longest-length states are dominant, and a clear oscillation period $T=2\pi/[\delta(L-1)]$ may be extracted, from which $\delta$ is obtained.

\end{appendices}
\end{document}